\documentclass[11pt]{article}
\usepackage[margin=1.2in]{geometry}
\usepackage{amsmath,amssymb,amsfonts,amsthm}
\usepackage{parskip}
\usepackage{xspace}
\usepackage{xcolor}       
\usepackage{amsmath,amssymb,amsthm}
\usepackage{wrapfig}
\usepackage{graphicx}
\usepackage{graphbox}
\usepackage{algpseudocode}
\usepackage{algorithm}
\usepackage{authblk}
\usepackage{subcaption}
\usepackage{float}
\usepackage{booktabs}
\usepackage{listings}
\usepackage{dirtree}
\graphicspath{{figures/}}

\definecolor{codegreen}{rgb}{0,0.9,0}
\definecolor{codegray}{rgb}{0.5,0.5,0.5}
\definecolor{codepurple}{rgb}{0.58,0,0.82}
\definecolor{backcolour}{rgb}{0.1,0.1,0.1} 
\definecolor{codecolor}{rgb}{0,0.7,0} 

\lstdefinestyle{mystyle}{
    backgroundcolor=\color{backcolour},
    commentstyle=\color{codecolor}, 
    keywordstyle=\color{magenta},
    numberstyle=\tiny\color{codegray},
    stringstyle=\color{codepurple},
    basicstyle=\ttfamily\footnotesize\color{codecolor}, 
    breakatwhitespace=false,
    breaklines=true,
    captionpos=b,
    keepspaces=true,
    numbers=left,
    numbersep=5pt,
    showspaces=false,
    showstringspaces=false,
    showtabs=false,
    tabsize=2
}

\usepackage[numbers,compress]{natbib}
\usepackage[pagebackref=true]{hyperref}
\usepackage{xcolor}
\definecolor{myblue}{rgb}{0.36, 0.54, 0.66}
\hypersetup{
	colorlinks=true,
	linkcolor=magenta,
	citecolor=myblue
}

\renewcommand{\geq}{\geqslant}
\renewcommand{\leq}{\leqslant}
\renewcommand{\subset}{\subseteq}

\newtheorem{proposition}{Proposition}
\theoremstyle{definition}

\newcommand\E{\mathbb{E}}
\newcommand{\PP}{\mathbb{P}}
\newcommand{\QQ}{\mathbb{Q}}

\usepackage{tocbasic}
\DeclareTOCStyleEntry[
  beforeskip=.05em plus .6pt,
  pagenumberformat=\textbf
]{tocline}{section}

\newif\ifarxiv
\arxivtrue

\title{Statistical Inference Under Constrained Selection Bias}

\author[1]{Santiago Cortes-Gomez}
\author[2]{Mateo Dulce}
\author[3]{Carlos Patino}
\author[1]{Bryan Wilder}

\affil[1]{Department of Machine Learning, Carnegie Mellon University}
\affil[2]{Department of Statistics, Carnegie Mellon University}
\affil[3]{Factored AI}
\affil[1]{\texttt{\{scortesg, bwilder\}@cs.cmu.edu}}
\affil[2]{\texttt{mdulceru@andrew.cmu.edu}}

\begin{document}

    \maketitle
    \begin{abstract}

  Large-scale datasets are increasingly being used to inform decision making. While this effort aims to ground policy in real-world evidence, challenges have arisen as selection bias and other forms of distribution shifts often plague observational data. Previous attempts to provide robust inference have given guarantees depending on a user-specified \textit{amount} of possible distribution shift (e.g., the maximum KL divergence between the observed and target distributions). However, decision makers will often have additional knowledge about the target distribution which constrains the \textit{kind} of possible shifts. To leverage such information, we propose a framework that enables statistical inference in the presence of selection bias which obeys user-specified constraints in the form of functions whose expectation is known under the target distribution. The output is high-probability bounds on the value of an estimand for the target distribution. Hence, our method leverages domain knowledge in order to partially identify a wide class of estimands. We analyze the computational and statistical properties of methods to estimate these bounds and show that our method can produce informative bounds on a variety of simulated and semisynthetic tasks, as well as in a real-world use case.
  
\end{abstract}
    
\section{Introduction}





Decision-makers in the public and private sectors increasingly seek to use machine learning or statistical models built on top of large-scale datasets in order to inform policy, operational decisions, individualized treatment rules, and more. However, these administrative datasets are typically purely observational, meaning that they are not carefully designed to sample from a true distribution of interest. Consequently, such efforts have been hindered by sampling biases and other distributional shifts between the observed data and the target distribution. Such selection biases have presented severe issues for past algorithmic systems  \cite{chouldechova2018case,gianfrancesco2018potential}, policy analysis \cite{harron2017challenges,knox2020administrative}, epidemiological studies \cite{haneuse2016general,jensen2015enrollment}, among others.

As a motivating example, decision-makers in public health may wish to estimate the risk factors contributing to adverse outcomes related to COVID-19, e.g., hospitalizations, in order to target preventative interventions. Nevertheless, they usually only have data from people who engaged with some specific services in the past such as insurance claims or patients who visited the health system. This scenario poses the difficulty that some demographic groups will be underrepresented in the data, influenced by some latent factors (for instance, high healthcare costs) and the outcome itself. It is a known phenomenon that individuals with worse healthcare access will seek care primarily when facing severe diseases, thus increasing observed hospitalizations for this group. Hence, creating a problem akin to confounding when the relationship between these two quantities is estimated. 

We consider the task of accurately estimating some functional of the ground-truth distribution using samples from an observed, potentially shifted, distribution. For instance, our goal might be to estimate the expectation of a measured covariate or the treatment effect from an intervention. In general, such inference is intractable without assumptions on the relationship between the observed and target distributions. The simplest setting is where we can directly observe some samples from the target distribution. However, we are interested in the scenarios where this is not the case, for example, the scenario introduced in the previous paragraph; given that the data is conditioned among other things on the outcome, it presents the insidious problem that it will never be representative of the population as whole. Absent target-distribution samples, other frameworks such as those related to distributionally robust optimization (DRO) \cite{bertsimas2022distributionally,gupta2021s,yadlowsky2022bounds,rothenhausler2022distributionally} or sensitivity analysis \cite{ding2016sensitivity,robins2000sensitivity,tan2006distributional} allow users to specify the total \textit{magnitude} of distribution shift allowed. Such magnitude is usually modeled by imposing a limit on the maximum distance (e.g., KL or $\chi^2$ divergence) between the target distribution and the observed one. However, since the target distribution (or samples of it) is unavailable, this assumption cannot be empirically verified, and often cannot even be set in an informed manner. Our goal is to provide provable guarantees \textit{without introducing untestable assumptions on the magnitude of shift}, by using only observable quantities to set bounds on the possible difference in distributions.

Intuitively, this may be possible because decision makers often have aggregate knowledge about the target distribution. For instance, a policymaker may know, via census data, the distribution of demographic characteristics such as age, race, or income from the population as a whole. Moreover, in the public health setting, serosurveys may provide ground-truth estimates of exposure to infectious diseases in specific locations or population groups \cite{havers2020seroprevalence}. This aggregate information is typically not sufficient by itself for the original task because it fails to measure the key outcome or covariates of interest  (e.g., knowing the demographic distribution of a population is by itself unhelpful for estimating a patient's risk of hospitalization). However, it imposes constraints on the nature of the distribution shift between the observed distribution and the underlying population: any valid shift must respect these known quantities. Concretely, our contributions are as follows:

\begin{itemize}
    \item  We introduce a framework that allows user-specified constraints on distribution shift via known expectations. Our framework incorporates such external information into an optimization program whose value gives valid max/min bounds on a statistic of interest using samples from an observed distribution. As a result, we are able to provide estimates by adding more observables in lieu of assumptions that are not empirically testable (e.g., the KL divergence between the target and sample distributions). We provide conditions under which this optimization problem is convex and hence, efficiently solvable.
    
    \item We analyze the statistical properties of estimating these bounds using a sample average approximation for the optimization problem. We show that our estimators are asymptotically normal, allowing us to provide valid confidence intervals.
    
    \item We extend our framework to accommodate estimands without a closed form (e.g., a regression coefficient or an estimated model parameter), provide statistical guarantees for this setting, and propose computational approaches to solve the resulting optimization problem.  
    
    \item We perform experiments on synthetic and semi-synthetic data to test our methods. The results empirically confirm that our framework provides valid bounds for the target estimand and allows effective use of domain knowledge: incorporation of more informative constraints produces tighter bounds, which can be strongly informative about the estimand of interest. All the code we used for our experiments is available at \url{hidden}.
    
    \item We showcase a real-world case of study of assessing disparities in COVID-19 disease burden using a large-scale insurance claims dataset where our method allows us to produce policy-relevant results under the presence of selection bias.
\end{itemize}

\paragraph{Additional related work.} Our work is broadly related to the literature on partial identification, which spans statistics, economics,  epidemiology, and computer science \cite{ho2015partial, manski2003partial, molinari2020microeconometrics, tamer2010partial}. Interest in partial identification has grown recently in the machine learning community, particularly in the causal inference setting. For example, \cite{hu2021generative,balazadeh2022partial} consider partial identification of treatment effects for a known causal graph, extending the classic framework of \cite{balke1997bounds} to incorporate generative models. \cite{guo2022partial} consider the setting where covariates are subject to a user-specified level of noise. Perhaps closest to our setting is \cite{padh2023stochastic}, who consider methods for incorporating domain knowledge into partial identification problems. However, in their setting, domain knowledge takes the form of functional form restrictions for the treatment effect (e.g., smoothness or number of inflection points), rather than constraints on shift between observed and target distributions. Our work differs from the existing ML literature both in that we are concerned with inference problems broadly (not restricted to treatment effects), and in that we provide a means to impose externally-known constraints on shifts such as selection bias. Related issues have recently been considered in the epidemiology and biostatistics literature, motivated by the growing use of biobank-style datasets with known selection biases  \cite{horowitz2022inference,tudball2019sample}. Our work differs in that we explicitly consider the algorithmic properties of computing the resulting bounds, and in our analysis of estimators which themselves require fitting a model.  

\section{Problem formulation}

Let $X$ be a random variable over a space $\mathcal{X}$, with distribution $\PP$ in a population of interest. Our goal is to estimate a real-valued functional $f(\PP)$. One prominent example is estimating the expectation $f(\PP) = \E_{X \sim \PP}[h(X)]$ for some function $h$, but we will consider a range of examples for $f$. If we observe iid samples from $\PP$, estimating $f(\PP)$ is a standard inference task. For instance, we may use the plug-in estimate $f(\hat{\PP}_N)$, (where $\hat{\PP}_N$ is the empirical distribution) or any number of other strategies. However, we consider a setting where we instead observe samples drawn iid from a different distribution $\QQ$. For illustration purposes, recall the example mentioned in the introduction: policymakers are interested in information relative to the population as a whole ($f(\PP)$) but they only have access to observational data from the hospitals or insurance claims, i.e., a sample from $\QQ$.

 We assume that both $\PP$ and $\QQ$ have densities $p$ and $q$, respectively, and that $q(X) > 0$ whenever $p(X) > 0$. More formally, we require that $\PP$ is absolutely continuous with respect to $\QQ$ so that the ground truth density ratio $\theta_0(X) = \frac{p(X)}{q(X)}$ is well defined. Intuitively, inference on $\PP$ is impossible if some portions of the distribution can never be observed. Accordingly, similar overlap assumptions are near-universal in causal inference \cite{imbens2004nonparametric}, domain adaptation \cite{byrd2019effect}, selection bias \cite{rosenbaum1983central}, distributionally robust optimization \cite{namkoong2016stochastic}, etc. We expect this assumption to be satisfied in many application domains because selection bias is not strong enough to make some observations literally impossible. Our running COVID-19 example illustrates this: a patient with any covariates $X$ can in principle seek care (and some such patients do), even if they have worse access than patients with a different set of covariates $X'$.
 
With slight abuse of notation, for any $\PP$, we let $\theta \PP$ denote a distribution with density $\theta(X)p(X)$ so that $\theta_0\QQ = \PP$. If $\theta_0$ were known, estimating $f(\PP)$ using samples from $\QQ$ would be easily accomplished using standard importance sampling methods: in the case of estimating an expectation, we have that $\E_{X \sim \PP}[h(X)] = \E_{X \sim \QQ}[\theta_0(X)h(X)]$. However, our focus is on the setting where $\theta_0$ is \textit{not} known, and we don't have access to samples from $\PP$ from which to estimate it. 

Without any information about the relationship between $\PP$ and $\QQ$ this is a hopeless task. However, in many settings of importance \textit{some} information is available, in the form of auxiliary functions whose true expectation under $\PP$ is known. In policy settings, this may come from census data, or well-design population surveys that estimate the ground-truth prevalence of specific health conditions (c.f.\ \cite{johnson2014national,havers2020seroprevalence,murray2013measuring}). Concretely, we are able to observe $\E_{\PP}[g_j(X)]$ for a collection of functions $g_1...g_m$. Thus, we know that $\theta_0$ must satisfy 
\begin{align*}
      \E_{\QQ}[\theta_0(X) g_j(X)] = \E_{\PP}[g_j(X)] \quad j = 1, \dots, m.
\end{align*}
Let $\Theta$ to be the set of density ratios $\theta$ which satisfy the above constraints, all of which are consistent with our observations. In general, $\Theta$ will not be a singleton and so $f(\PP)$ will not be point-identified. However, it is \textit{partially} identified with the following bounds:

\begin{proposition}  
\label{prop:identification}
$$f(\PP) \in \left(\min_{\theta \in \Theta} f(\theta \QQ), ~~ \max_{\theta\in \Theta} f(\theta \QQ) \right),$$
and these bounds are tight.
\end{proposition}

In the following, we provide computationally and statistically efficient methods for estimating these upper and lower bounds, each of which are defined via an optimization problem over $\theta$. $\Theta$ encapsulates the constraints on potential distribution shifts that are known in a particular domain, allowing an analyst to translate additional domain knowledge into tighter identification of the estimand. Intuitively, such identification will be close to point-wise when the constraints are informative enough about the behaviour of the estimand of interest under the true distribution. Our aim is to use the observed sample $X_1,\dots,X_N$ drawn iid from $\QQ$ to estimate the value of the optimization problems defining our lower and upper bounds. Note that in order to do so, we have to tackle two different types of uncertainty. First, the uncertainty derived from partial identification where even the population level quantities $ \min_{\theta \in \Theta}/\max_{\theta \in \Theta} f(\theta\QQ)$ do not exactly identify $f(\PP)$ even if $\QQ$ were known exactly. Second, finite-sample uncertainty from the estimation of the population level quantity $f(\theta\QQ)$ using a sample $X_1,\dots,X_N$: the statistical uncertainty from taking  $f(\theta\hat{\QQ}_N)$ as an approximation of $f(\theta\QQ)$. We will provide confidence bounds for $f(\PP)$ which account for both partial identification and finite-sample uncertainty in estimation. Throughout, we will assume for convenience that the optimization problems defining each side of the bound have a unique solution (if this is not satisfied, we could, e.g., modify the objective function to select a minimum-norm solution).

\section{Methods}

We now proceed to estimate the aforementioned bounds. Proposition \ref{prop:identification} reduces the problem of computing such bounds to solving an optimization program. Throughout, we consider the problem of estimating the lower bound (i.e., solving the minimization problem), as the upper bound is symmetric. To start, we impose no restrictions on $\theta(X)$ by representing $\theta(X)$ as an arbitrary function. This yields the following optimization problem:
\begin{equation}
\label{pop_problem1}
\begin{split}
    &\min_\theta f(\theta \QQ) \\
    &\theta(X) \geq 0 \quad \forall X,\\
    &\E_{X \sim  \QQ} [\theta(X) g_j(X)] = \E_{\PP}[g_j(X)] \quad j = 1,\dots,m. 
\end{split}
\end{equation}
where the constraint $\theta(X) \geq 0$ ensures that $\theta$ is a valid density ratio. However, neither $f(\theta \QQ)$ nor the population level restrictions can be observed directly. Nevertheless, having access to a sample from $\QQ$, it is only natural to use plug-in estimators to estimate the unavailable quantities from the previous program. This yields the sample optimization problem:
\begin{equation}
\label{sample_problem1}
\begin{split}
     &\min_\theta f(\theta \hat{\QQ}_N) \\
    &\theta(X_i) \geq 0 \quad \forall i = 1,\dots,N,\\
    &\frac{1}{N} \sum_{i  =1}^N \theta(X_i) g_j(X_i) = \E_{\PP}[g_j(X)] \quad j = 1,\dots,m
\end{split}
\end{equation}
where $\hat{\QQ}_N$ is the empirical distribution of $\QQ$. We refer to the first problem \eqref{pop_problem1} as the \textbf{population optimization problem} ($\min_\theta f(\theta \QQ)$) and to the second one \eqref{sample_problem1} as the \textbf{plug-in approximation} ($\min_\theta f(\theta \hat{\QQ}_N)$). 

Let $\nu$ be the optimal value of the population problem and $\hat{\nu}_N$ the optimal value of the plug-in approximation problem. Suppose $\sqrt{n}(\hat{\nu}_N - \nu)$ converges in distribution to a normally distributed random variable, such that, it has mean zero and variance that can be estimated from the data. If this is indeed the case, then the standard procedure to derive confidence intervals for $\nu$ can be applied. Furthermore, as corollary of Proposition \ref{prop:identification}, we can produce a high probability lower bound for $f(\PP)$. As the maximization problem is symmetric, we can use the same argument to upper bound $f(\PP)$. 

We now turn to analyze several scenarios where this recipe can be applied. We use the theory of \textit{sample average approximation} in optimization \cite{shapiro1991asymptotic, shapiro2021lectures} to describe the conditions to guarantee, both convergence and asymptotic normality, for the statistic $\sqrt{n}(\hat{\nu}_N - \nu)$. However, before developing this mathematical scaffolding, it is necessary to make one more assumption: we will assume a, differentiable, finite-dimensional parametrization of $\theta(X)$, denoted by $\theta_{\alpha}(X)$, for some $\alpha \in \mathbb{R}^d$. This class is still quite expressive; e.g., a standard basis for smooth functions (e.g., a set of polynomial basis functions) allows us to represent any smooth function in this framework, or the analyst could use a highly expressive class such as neural networks \cite{hornik1989multilayer}.

\subsection{Convex estimands}

We start with the case where $f(\theta_{\alpha} \QQ)$ is a convex function of $\alpha$. The most prominent case where this holds is when $f$ is the expectation of some function $h$, in which case $f(\theta_{\alpha} \hat{\QQ}_N) = \frac{1}{N}\sum_{i=1}^N \theta_{\alpha}(X_i) h(X_i)$ is linear in $\alpha$ and will be convex in $\alpha$ when $\alpha \mapsto \theta_\alpha$ is convex. Another example is when $f$ is a conditional expectation, conditioned on some event $X \in C$. Then, we have
\begin{align*}
    f(\theta_{\alpha} \QQ) &= E_{X \sim \theta_{\alpha} \QQ}[h(X) | X \in C], \\
    f(\theta_{\alpha} \hat{\QQ}_N) &= \frac{\sum_{i = 1}^N 1[X_i \in C] \cdot \theta_{\alpha}(X_i) h(X_i)}{\sum_{i = 1}^N \theta_{\alpha}(X_i) 1[X_i \in C]}. 
\end{align*}
The plug-in approximation of $f$ is no longer linear in $\theta_{\alpha}$ because $\theta_{\alpha}$ determines both the numerator and the denominator. However, it can be reformulated as a linear function using the standard Charnes-Cooper transformation \cite{zionts1968programming} and thus can be computed by the means of a linear program. It is worth pointing out that, being linear programs, the previous two examples can be efficiently computed.

When $f$ is a convex function of $\alpha$, we can use standard results in sample average approximation programs \cite{shapiro2021lectures} to show asymptotic normality. Let $\lambda_j$ be the dual variable associated with constraint $j$, and $\lambda^*_j$ be the optimal value of $\lambda_j$ in the population problem. Similarly, let $\alpha^*$ be the population-optimal value of $\theta_{\alpha}$. Then, we have 
\begin{proposition}
    Let $f(\theta_{\alpha}\hat{\QQ}_N)$ be convex in $\alpha$. Moreover, assume $\alpha \in S$ for $S$ compact and that the Slater's condition holds. Then, $\sqrt{N}(\nu - \hat{\nu}_N) \to \mathcal{N}(0, \sigma^2)$ and convergence is in distribution. In particular, if $f(\theta_{\alpha}\QQ) = \mathbb{E}_{\QQ}[h(x)\theta_{\alpha}]$, then $\sigma^2 = \text{Var}[\theta_{\alpha^*}(X)h(X) + \sum_{ j =1}^M \lambda^*_j (\theta_{\alpha^*}(X)g_j(X) -  \E_{\PP}[g_j(X)])]$.
\end{proposition}

We can then produce confidence intervals by estimating this variance via the sample estimates $\hat{\theta}_N$ and $\hat{\lambda}_{N, j}$ (see e.g. Schapiro et al. \cite{shapiro2021lectures} Eqs. 5.183 and 5.172). In order to simplify estimation of this variance, it may be recommended to use a form of sample splitting (at the expense of a slower convergence rate), where disjoint sets of samples are used to estimate the objective function and each constraint in the sample problem. This eliminates the covariance terms in the expression for $\sigma$ above. Alternatively, if computational power is not a constraint, it may be simpler in practice to estimate the variance using the bootstrap. 

\subsection{General estimands}

When the estimand $f(\theta_{\alpha}\QQ)$ is a non-convex function of $\alpha$, obtaining provably optimal solutions for the plug-in approximation problem is in general not feasible. However, we can still obtain locally optimal solutions (as is common for other partial identification settings \cite{padh2023stochastic,horowitz2022inference,hu2021generative,balazadeh2022partial}), although the statistical properties of the plug-in estimator $f(\theta_{\alpha} \hat{\QQ}_N)$ may become more complex, as we illustrate in the two following examples. 

\textbf{Example 1: Average treatment effect estimation}. Consider the setting where $\PP$ is a distribution over tuples $(X, Y, A)$, where $X$ is a covariate vector, $Y$ is an outcome, and $A$ is a binary treatment indicator variable. For simplicity, we consider the case where the outcome $Y$ is also binary. Researchers are often interested in estimating the average treatment effect. Under standard identifying assumptions (most prominently that $A \perp Y^{A=a} | X$), this is done by the means of the estimand:
\begin{align*}
    f(\PP) = \int_X \left[p(Y|A = 1, X) - p(Y|A= 0, X)\right]d\PP(X) 
\end{align*}
Now, consider the setting where we observe samples from a distribution different than $\PP$, resulting in a density ratio $\theta(X, Y, A)$. Computing the appropriate marginals and conditionals of $\theta(X, Y, A)p(X, Y, A)$, and substituting these for $p$ in the above expression, gives an objective that is non-convex in terms of $\theta$, but which is nonetheless differentiable, enabling gradient-based optimization. However, we will still have to estimate the nuisance functions $p(Y = 1|A=a, X)$ and the statistical properties of the resulting bounds will depend on how these are estimated.

\textbf{Example 2: Coefficients of parametric models}. Suppose that a researcher is interested in interpreting the estimated coefficient of a parametric model, e.g., linear or generalized linear models as commonly used in a variety of applied settings. For example, an applied researcher may estimate the odds ratio for an outcome given some exposure using a logistic model and wishes to obtain bounds on this parametric odds ratio under potential selection bias or other distribution shifts. 

\textbf{Bounding $M$-estimators:}
To provide one way of addressing these (and other) examples, we consider the general challenge of partially identifying quantities produced by an $M$-estimator. $M$-estimators are those which estimate a parameter $\beta$ via minimizing the expected value of a function $m$, i.e., $f(\PP) = \arg\min_\beta \E_{X \sim \PP}[m(X, \beta)]$, widely used across many areas of statistics and machine learning \cite{van2000asymptotic}. One prominent example is when $m$ is the negative log likelihood, resulting in a maximum likelihood estimate of $\beta$. We are interested in producing bounds for some real-valued function $h$ of $\beta$. For example, $h$ may be the value of a single coordinate of $\beta$ if we are interested in bounding a specific model coefficient that will be interpreted, or $h$ may be the treatment effect functional described above. This results in the optimization problems: 
\begin{align*}
    &\min_\alpha h(\beta(\theta_{\alpha})) \\
    &\beta(\theta_{\alpha}) = \arg \min_\beta \E\left[\theta_{\alpha}(X) m(X, \beta)\right] \\
     &\theta_{\alpha}(X) \geq 0 \\
    &\E_{X \sim  \QQ}[\theta_{\alpha} g_j(X)] = \E_{\PP}[g_j(X)] \quad j = 1, \dots, m\\ 
    \\
    \hline
    \\
    &\min_\alpha h(\hat{\beta}_N(\theta_{\alpha}))\\
    &\hat{\beta}_N(\theta) = \arg \min_\beta \frac{1}{N}\sum_{i=1}^N\theta_{\alpha}(X_i) m(X_i, \beta)\\
    &\theta_{\alpha}(X_i) \geq 0 \quad i=1,\dots, N \\
    &\frac{1}{N} \sum_{i  =1}^N \theta_{\alpha}(X_i) g_j(X_i) =  \E_{\PP}[g_j(X)] \quad j = 1,\dots,m.
\end{align*}

Since $h(\beta(\theta_{\alpha}))$ is not convex in general, Slater's condition is not a strong enough regularity condition to guarantee asymptotic normality of the estimated optimal value. Therefore, we require to impose a more general constraint qualification. Specifically, we will assume the standard \textit{Mangasarian-Fromovitz Constraint Qualification} \cite{kyparisis1985uniqueness} for the non-convex case. This regularity condition is sufficient to ensure that each minimizer of the population problem has a unique $\lambda^*_j$ associated in the dual.  Hence, under the MFCQ regularity condition and if $\Sigma^*$ is the covariance matrix of the $M$-estimator at the optimal $\alpha$ we prove the following asymptotic normality result:
\begin{proposition}
    Assume that $m$ is twice differentiable and locally convex in $\beta$. Assume as well that $||\nabla_\alpha \theta_\alpha(x)||_2 \leq \kappa_\theta(x)$, $||\nabla_\beta m(x, \beta)||_2 \leq \kappa_m(x)$,  $||\nabla_\beta h(\beta)||_2 \leq \kappa_h$ . Then, if $\alpha$ lies in a compact set $S$, $h(\alpha)$ and $g_i(x)$ are differentiable, then $\sqrt{N}(\hat{\nu}_N - \nu) \to N(0, \sigma^2)$. Using sample splitting, the variance can be approximated by $\nabla h(\beta^*(\alpha^*))^T \Sigma^* \nabla h(\beta^*(\alpha^*)) + Var(\sum_{i=j}^m\lambda^*_j \theta_{\alpha^*}(x)g_j(x))$.
\end{proposition}


The proof of Proposition 3 requires tools from empirical process theory in order to study the asymptotic properties of the optimal value of the plug-in approximation problem. To sketch the main idea, let $Y_n$ be the vector of functions for the plug-in approximation problem and $\mu$ for the population problem (where the first coordinate in each is the objective function and the rest are the constraints). Let $\psi$ be the function $\psi(\mu) = \min_{\theta_{\alpha} \in \Theta} f(\theta_{\alpha}\QQ)$.  If the empirical process $\sqrt{n}(Y_n - \mu)$ converges in distribution to an object, then a functional version of the delta method can be applied to obtain our result. This strategy requires two conditions. First, that a limiting object exists (in distribution) for the random vector $\sqrt{n}(Y_n - \mu)$, which we accomplish by showing that the empirical process belongs to a Donsker class. Second, that such a limiting object, with specified variance, still exists after applying the function $\psi$. We show this, subject to an appropriate constraint qualification, by applying a generalized version of the continuous mapping theorem (as in \cite{shapiro1991asymptotic}). The whole proof and explicit expressions for $\sigma^2$ can be found on the Appendix.

\subsection{Computational approach}

As long as $f(\theta_{\alpha})$ has well defined subgradients, the plug-in approximation problem can be solved efficiently by projected gradient descent. Even for programs that include an $M$-estimator, this is easily accomplished in autodifferentiation frameworks where we can use differentiable optimization \cite{amos2017optnet,agrawal2019differentiable} or meta-learning style methods \cite{bertinettometa,finn2017model,lee2019meta} to implement the $\arg\min$ defining $\beta$ in a manner which supports automatic backpropagation. In general, we can only obtain locally (instead of globally) optimal solutions for non-convex problems. However, we observe experimentally that the values obtained are nearly identical across many random restarts of the optimization, suggesting good empirical performance.

\section{Experiments}

We conduct experiments to show how our method allows users to specify domain knowledge in order to obtain informative bounds on the estimand of interest. We simulate inference for a range of different $f(\PP)$ across various scenarios by testing our method with different choices of constraints and datasets. In addition, we present a real-world use case where we were able to produce policy-relevant conclusions using our method.

In each experiment, we start with samples from a ground truth distribution $\PP$ and then simulate the observed distribution $\QQ$ using sampling probabilities which depend on the covariates (i.e., simulating selection bias). This ensures that the ground-truth value $f(\PP)$ is known (to high precision), allowing us to verify if our bounds contain the true value. We consider two classes of estimands. First, estimating the conditional mean $\mathbb{E}_{\PP}[Y | A=1]$ for an outcome $Y$ and covariate $A$. Second, estimating the coefficient of a linear regression model as an example of the $m$-estimation setting from above. 

We show how the amount of specified domain knowledge can result in tighter bounds along two axes. First, we vary the parametric form for $\theta$, ranging from an arbitrary function of $X$ (i.e., a separate parameter for each discrete covariate stratum) to one that imposes specific assumptions (e.g., separability across specific groups of covariates, or that the covariates which the selection probabilities depend on are known).  Second, we vary the number and informativeness of the constraints $\{g_j\}$ used to form the set $\Theta$; modeling the ability of users to impose an increasing degree of constraints on possible distribution shifts. The Appendix shows experiments with additional kinds of constraints, e.g., on the sign of the covariance between pairs of variables. 

The closest baselines to compare our methods against are sensitivity analyses based on distributionally robust optimization (DRO). However, DRO requires the distance between distributions under some divergence (the unobservable parameter $\rho$) as input. Conversely, our method requires observable aggregate data to provide partial identification, making a head-to-head comparison difficult. In order to provide baselines for comparison, we implement DRO on two artificial settings where the maximum distance $\rho$ is provided. The first baseline (\textit{omniscient}) optimistically assumes that the $\chi^2$ divergence ($\rho$) between $\hat{\QQ}_n$ and  $\PP$ is known. This gives an "unfair advantage" to DRO because not only is a valid value of $\rho$ provided but, furthermore, it is the smallest of such feasible values (which is not actually observable when $\PP$ is unknown). Hence, we also propose a second scenario (\textit{observable}) where the tightest possible value of $\rho$ is inferred from the same data used by our method. In this scenario $\rho$ is estimated by solving for $\max_{\theta \in \Theta} D_{\chi^2}(\theta\hat{\QQ}_n, \hat{\QQ}_n)$. Thus, the solution of the program will be the radius of the smallest ball centered at $\hat{\QQ}_n$ that contains all distributions that satisfy all the observable constraints. Of course the previous program is not convex in $\theta$; hence we settle for the solution obtained from a saddle point obtained via gradient descent (which errs in favor of DRO by selecting a smaller value of $\rho$ than the one that may be justified by the data).

We now proceed to summarize how these experiments were instantiated in each setting, providing detailed information in the Appendix. Additionally, a guide on how to replicate each of these experiments is available in the code repository at \url{hidden}.

\textbf{Synthetic data experiments.} For the first set of experiments, we simulate a distribution over binary variables  $X = (Y,Y_2, A,X_1, X_2) \sim \PP$ used to evaluate previous causal inference methods \cite{kennedy2019estimating}. We add a selection bias scenario to this process by simulating an indicator variable $R \sim Ber(logit^{-1}(X_1 - X_2))$. The observed distribution $\QQ$ consists of those samples for which $R=1$. In this domain, we consider the task of estimating bounds for $\mathbb{E}_{\PP}[Y | A=1]$ setting restrictions on $\mathbb{E}_{\QQ}[\theta YX_2]$ (i.e., $\mathbb{E}_{\PP}[YX_2]$ is known).  A total of six experiments were run. In the first three experiments, we vary the functional form for $\theta$ while leaving the constraints defining $\Theta$ fixed. In the first experiment (\textit{unrestricted}), $\theta$ is an arbitrary function of $X$. In the second experiment (\textit{separable)}, we specify $\theta(X) := \theta_1(A) + \theta_2(X_1,X_2)$ where $\theta_1$ is an arbitrary function of $A$ and $\theta_2$ is an arbitrary function of $X_1$ and $X_2$. Finally, in the third experiment (\textit{targeted}), we fix $\theta(X) = \theta(X_1, X_2)$, i.e., $\theta$ is a function only of the variables which determine $R$, thus simulating the scenario where the variables driving selection bias are known (even if the exact selection probabilities are not).

For experiments four to six, we vary $\Theta$ instead by adding more informative constraints in each successive experiment. As our intention is to isolate the effect of adding more constraints, the parametric form of $\theta$ is set as flexible as possible, i.e., as an arbitrary function of $X$ (the \textit{unrestricted} experiment above). In experiment four, we constrain two of the four parameters of the joint distribution of $Y_2$ and $X_2$ via constraints on $\mathbb{E}_{\QQ}[\theta Y_2X_2]$ and $\mathbb{E}_{\QQ}[\theta Y_2(1-X_2)]$. In experiment five, we add constraints on the remaining components of the joint distribution $\mathbb{E}_{\QQ}[\theta (1 - Y_2)X_2]$ and $\mathbb{E}_{\QQ}[\theta (1 - Y_2)(1 - X_2)]$. Experiment six, add constraints on the outcome ($\mathbb{E}_{\PP}[YX_2]$) as well. Through  out, we will refer to these experimental setups as \textit{(partial) race + income}, \textit{(full) race + income} and \textit{race + income + outcome} respectively (the names were chosen to keep consistency with the semi-synthetic experiments).
\begin{figure*}
  \centering

  \includegraphics[scale=0.21]{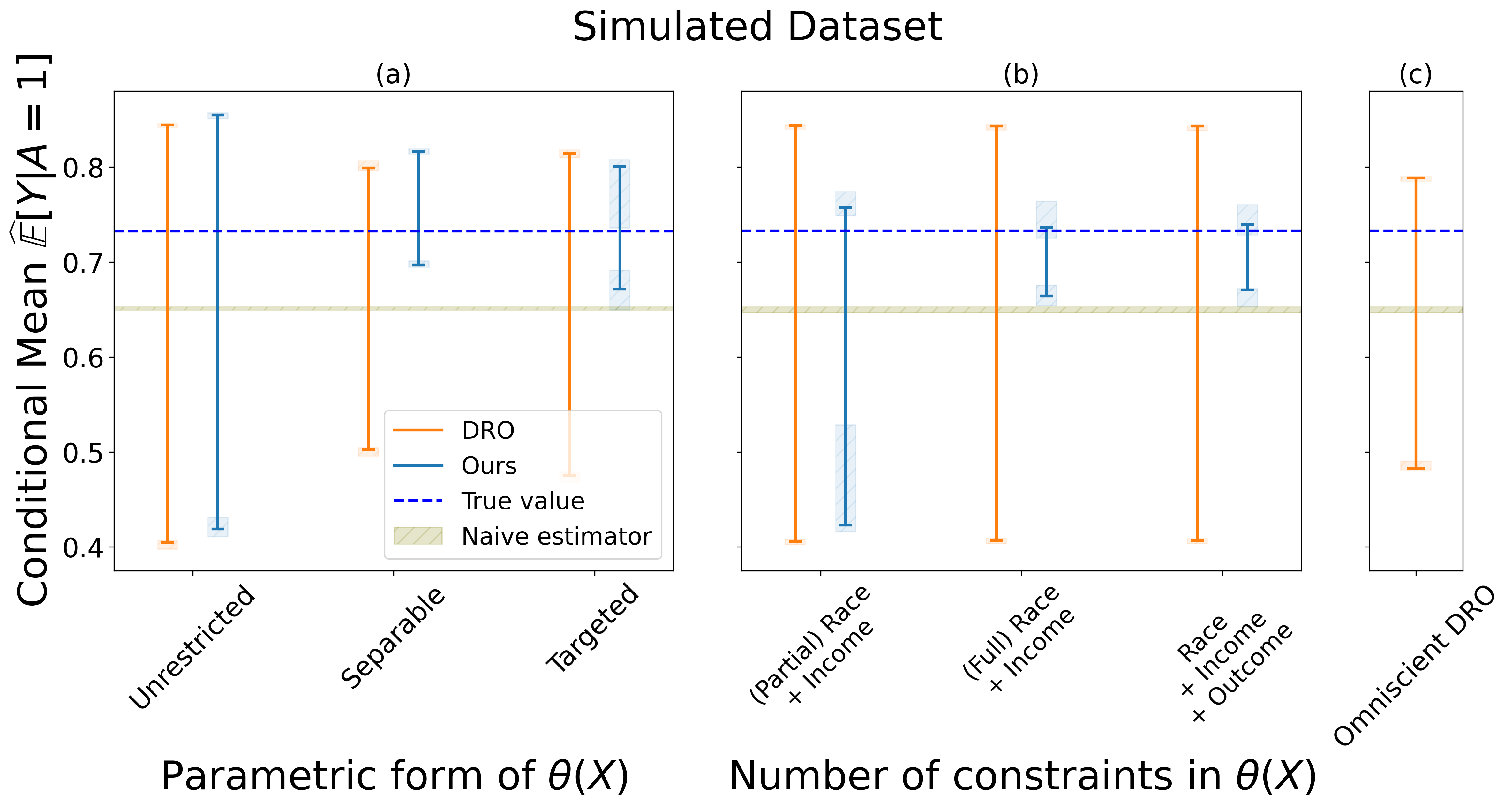}
  \includegraphics[scale=0.21]{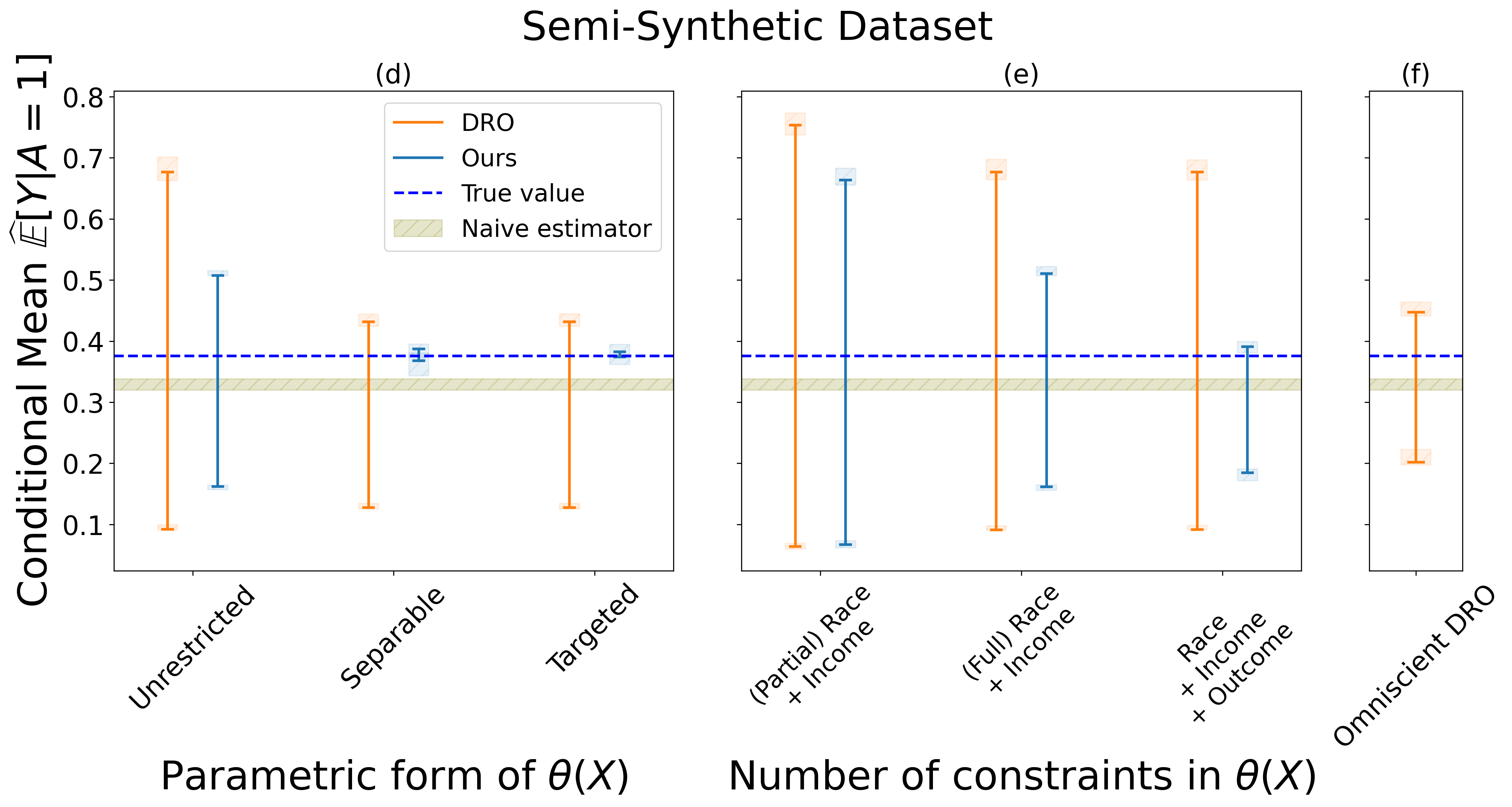}

  \caption{Partial identification of a conditional mean in simulated and semi-synthetic experiments using our proposed method vs. \textit{observed} DRO. The experiments include several parameterizations of $\theta(X)$ and different sets of constraints $\Theta$.}

    \label{bounds}
\end{figure*}

\begin{figure}
  \centering
  
  \includegraphics[scale=0.21]{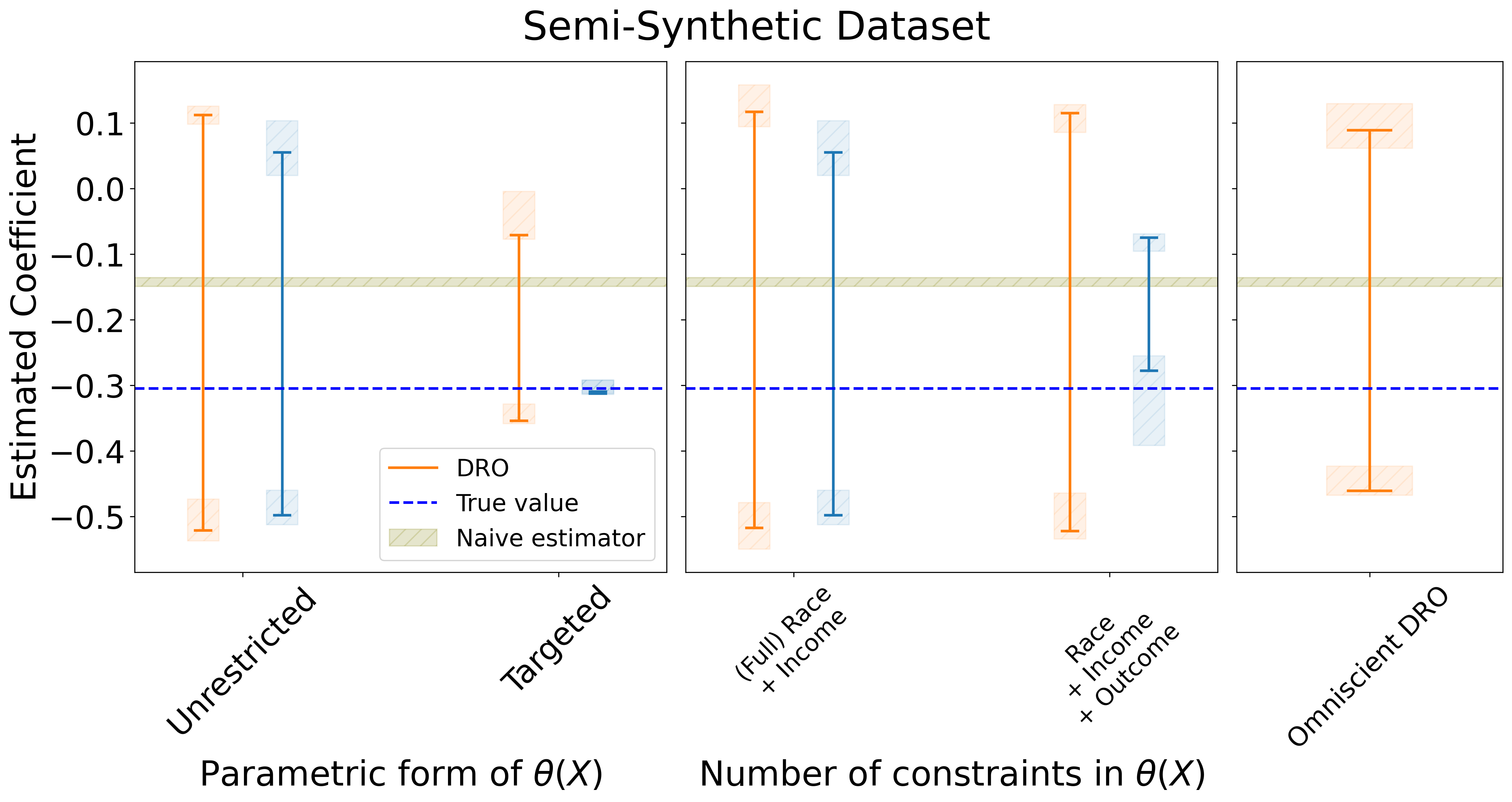}
  \caption{Estimated value of $\beta(\theta)$ with our proposed method for the semi-synthetic dataset. The first two images are results for different parameterizations of $\theta(X)$ and results for different sets of constraints $\Theta$, respectively. In both cases, incorporating more external information leads to narrower bounds.}

  \label{bounds_beta}
\end{figure}
\textbf{Semi-synthetic data experiments.} We use the Folkstables dataset \cite{ding2021retiring} which provides an interface for the data from the US Census American Community Survey from 2014 to 2019. Based on the ACSEmployment task suggested in \cite{ding2021retiring}, we consider a binary outcome variable $Y$ indicating whether or not a person was employed at the time of the survey. We are interested in predicting the female unemployment rate (e.g., motivated by a policymaker who seeks to track or intervene on gender employment gaps \cite{albanesi2018gender} \cite{mate2022field}). The variables driving the selection bias are citizenship and veteran status, as these variables can contribute to self-selection in a potential social services database. 

We first consider estimating $\E_{\PP}[Y|A = 1]$ where $A$ indicates the sex of an individual. We conduct six experiments that exactly parallel the construction of the synthetic experiments, with the variable for income level playing the role of $Y_2$ and race/ethnicity playing the role of $X_2$ (see Appendix for details).  This simulates a setting where information about other outcomes correlated with the variables of interest is leveraged for partial identification. In the last experiment \textit{race + income + outcome}, the additional constrain $\mathbb{E}_{\QQ}[ \theta YX_2]$ simulates a scenario where information on the outcome of interest is available with respect to a different demographic grouping than our desired estimand (race/ethnicity instead of gender).

We then consider estimating the coefficient of the indicator variable $A=1$ in a linear regression model of $Y$ on 7 covariates. This second setting is an example of the $M$-estimator framework from above. We run the experiments \textit{(partial) race + income}, \textit{(full) race + income} and \textit{race + income + outcome} using the settings of the $E_{\PP}[Y|A = 1]$ case. Details for each experiment are shown in the Appendix.

\textbf{Real world case study.}  We apply our method to a dataset of over five million de-identified medical claims from COVID-19 patients. Medical claims provide the services that patients received, their diagnosis, and demographic attributes. We use this data to study disparities in hospitalization risk for US COVID-19 patients across race/ethnicity groups. We provide bounds for the \textit{relative excess risk of hospitalization} for non-white patients after conditioning on clinically relevant covariates (a set of five comorbidities and age). Formally, our estimand is $f(\PP) = \sum_{x} [P(Y =1 |X=x, Race) - P(Y =1 |X=x, Race=White)]P(X=x | Race)/P(Y=1|Race=White, X)$, where $Y$ is whether a patient was hospitalized. As mentioned in the introduction, selection bias is particularly challenging for these data, because it may depend jointly on both the outcome and main covariate of interest. We model selection based on race, income, and hospitalization status to account for this potential interaction between disease severity and healthcare access in ascertainment. As constraints $g$, we use the total hospitalizations per racial group reported by the CDC and the total number of estimated infections per group from CDC serology studies. 

To further investigate the stability of our estimand, every experiment was run across 5 different random restarts of the projected gradient descent algorithm.

\begin{figure}
    \centering
    \includegraphics[scale=0.4]{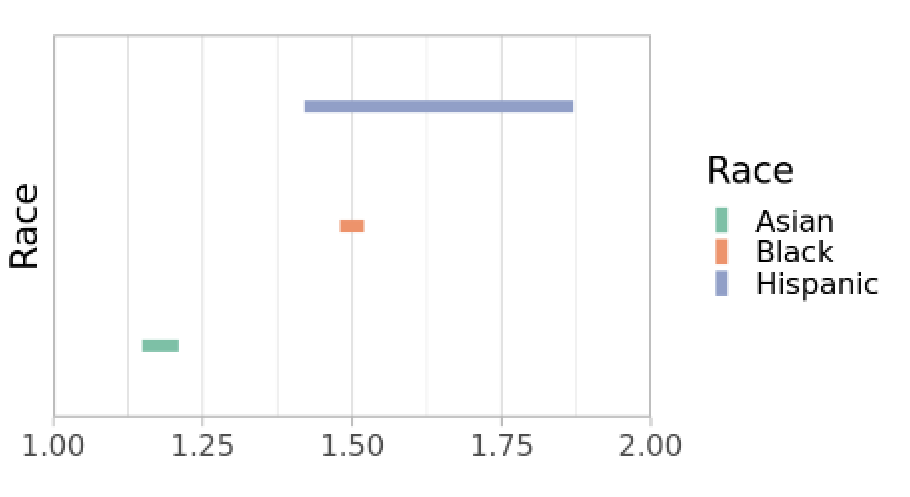}
    \caption{Estimated value of $f(\PP)$ by race. According to our method there is an increased risk of hospitalization for COVID-related emergencies for Asian, Black, and Hispanic populations when compared to the White population. For the Black and Hispanic population, our method of partial identification is close to a point-wise identification. }
  \label{real_world}
\end{figure}

\textbf{Results.} Figures \ref{bounds} and \ref{bounds_beta} show our main results. Each figure plots the bounds output by our method for each experiment described above. In each plot, we also show the true value of the estimand $f(\PP)$, the naive estimate using samples from $\QQ$ (without any attempt to account for selection bias) and the bounds produced by the two DRO baselines. In each plot (\ref{bounds} and \ref{bounds_beta}), accounting for the different results across multiple bootstrap replicates, there is a box plot at the endpoints of every bound as well as a band to represent $f(\QQ)$ . 

The results of our experiments are consistent across all six scenarios, showing how the estimated bounds vary depending on the constraints imposed and on the functional form assumed for $\theta$.  Concordant with the theory, the bounds always contain the true value of the estimand. Additionally, as expected, when more external information is available the bounds become narrower. When additional constraints are added, it is possible to obtain narrow bounds for the estimand that rule out the naive estimates. Our method Pareto-improves over the observed-data DRO baseline across all settings. Interestingly, in the majority of settings, our method also produces tighter intervals than the \textit{omniscient} DRO baseline. Thus, it can be highlighted the advantage of more nuanced descriptions of uncertainty: even in the case the true value of the divergence is known, moment constraints may provide a tighter bound than a divergence metric. We also find that in the \textit{targeted} experiment where the variables driving selection bias are known, our method often provides close to point identification, indicating that this is a particularly powerful form of domain knowledge when available. These results demonstrate that our framework allows users to translate common forms of domain knowledge into robust and informative statistical inferences. 

Figure \ref{real_world} summarizes the results from the real-world case study. We find that Black, Asian, and Hispanic patients have a higher risk of hospitalization compared to white patients. Importantly, our method produces informative bounds for all groups, which are \textbf{close to point identification} for Black and Asian patients while highlighting greater sensitivity to potential selection biases for Hispanic patients. Since the true underlying target distribution is unknown, we cannot compare our results to the `ground truth' in this domain; rather, this showcases that our methods produce informative and theoretically-guaranteed intervals for a policy-relevant question.

\newpage
\bibliographystyle{plainnat}
\bibliography{main}





    \newpage
    \appendix
    \section{Proofs}
\subsection{Proof proposition 1}
This holds by construction since $\PP = \theta \QQ$ for some $\theta \in \Theta$. To see that these bounds are tight, suppose \textit{wlog }that we have a conjectured lower bound $\ell > \min_{\theta
 \in \Theta} f(\theta \QQ)$. Then, there would be some $\theta$ consistent with all the constraints in $\Theta$ for which $f(\theta\QQ) < \ell$, implying that there is a target distribution $\PP$ consistent with all constraints for which $\ell$ is not a valid lower bound.

\subsection{Proof proposition 2}
    This follows directly from Theorem 5.11 of \cite{shapiro2021lectures}, since the objective and constraints are convex and we have assumed that the population problem has a unique solution. 

\subsection{Proof proposition 3}

We want to show that the empirical process $\sqrt{N}(\hat{\nu}_N - \nu)$ converges to a Gaussian process when $N\to \infty$, where $\widehat{\nu}_N$ is the value of the plug-in approximation program estimated from a sample of size $N$. To prove this, we will use the delta method in Theorem 2.1 of \cite{shapiro1991asymptotic} following the Hadamard differentiability of the outer problem $\min_{\alpha} f(\alpha, \xi)$. The later condition is shown to hold by invoking Theorem 3.6 in \cite{shapiro1991asymptotic}. However, to apply such theorem, the $\textit{Mangasarian-Fromovitz Constraint Qualification}$ has to hold, and the approximation vector has to have a limiting process in $C(S)$; the first requirement is satisfied by assumption, and the second is shown in the following.

\textit{Assumption 1.} $\theta(x) = \theta_\alpha(x)$ has some finite dimensional parameterization $\alpha \in S \subset R^d$. 

Consider the function $f(\alpha, \xi) = h(\beta^*(\theta_\alpha, \xi))$ where 
\begin{align*}
    \beta^*(\theta_\alpha, \xi) = \arg \min_{\beta} \E_\xi [\theta_\alpha(x_i) m(\beta, x_i)].
\end{align*}
and $\xi$ is the distribution that the expectation is taken with respect to, e.g., the realized empirical distribution on a set of samples $x_1, \dots, x_N$ or the limiting distribution $\QQ$.

\textit{Assumption 2.} The functions $f$ and $g_i$ are differentiable on $S$. 

\textit{Assumption 3.} $\beta^*$ is always a strict local minimizer for any $\theta$ and realization $\xi$ so that the Hessian matrix $H(\beta) = \nabla^2_\beta\left[\E_\xi[\theta_\alpha(x_i) m(\beta, x_i)]\right]$ is positive definite. In particular, we assume that the smallest eigenvalue satisfies $\lambda_{\text{min}}(H(\beta)) > \gamma_{\xi}$ for some $\gamma_\xi > 0$ which may depend on the realization of the randomness $\xi$, but not on $\theta$. 

\textit{Step 1.}
Our goal will be to show that $f$ is Lipschitz as a function of  $\alpha$ for any fixed $\xi$, which we will accomplish by showing that $||\nabla_\alpha f||$ is bounded.  We use $||\cdot||_2$ for matrices to denote the spectral norm, or operator 2-norm, and for vectors to denote the Euclidean norm. We assume that $||\nabla_\alpha \theta_\alpha(x)||_2$, $||\nabla_\beta m(x, \beta)||_2$, and $||\nabla_\beta h(\beta^*)||_2$ all satisfy bounds which in the case of the first two are allowed to depend on $x$:
\begin{align*}
    ||\nabla_\alpha \theta_\alpha(x)||_2 \leq \kappa_\theta(x), \quad ||\nabla_\beta m(x, \beta)||_2 \leq \kappa_m(x), \quad ||\nabla_\beta h(\beta)||_2 \leq \kappa_h.
\end{align*}

That is, they are each Lipschitz in their respective parameter for a fixed realization $x$, but with Lipschitz constant which is allowed to depend on $x$. While we present these assumptions in terms of the Euclidean norm for convenience, since we only aim to prove that the Lipschitz constant of $f$ is bounded (our results do not depend on its particular value), bounds on the gradients above in any norm suffice (using the equivalence of all norms in finite dimension). 

In what follows we fix $\xi$ and drop it from the notation. We first apply the chain rule to obtain that $\nabla_\alpha f = [D_\alpha \beta^*(\alpha)] \nabla_\beta h(\beta^*(\alpha)) $ where $D_\alpha \beta^*(\alpha)$ is the Jacobian of $\beta^*$ with respect to $\alpha$. To calculate this term, we apply the implicit function theorem to the first-order optimality condition
\begin{align*}
    \nabla_\beta \E[\theta_\alpha(x)m(\beta, x)] = 0
\end{align*}
and obtain
\begin{align*}
    D_\alpha \beta^*(\alpha) = H(\beta^*)^{-1} \nabla_\alpha \nabla_\beta\E[\theta_\alpha(x) m(\beta^*, x)]
\end{align*}
where we remark that $H$ is guaranteed to be invertible by our assumption that it is positive definite in a neighborhood of $\beta^*$ and hence the implicit function theorem applies.

 Next, note that 
\begin{align*}
    \nabla_\alpha \nabla_\beta\E[\theta_\alpha(x) m(\beta^*, x)] = \E[\nabla_\alpha \theta_\alpha(x) [\nabla_\beta m(x, \beta)]^T]
\end{align*}
and
\begin{align*}
    ||\E[\nabla_\alpha \theta_\alpha(x) [\nabla_\beta m(x, \beta)]^T]||_2 
    \leq \E[||\nabla_\alpha \theta_\alpha(x) [\nabla_\beta m(x, \beta)]^T||_2]
\end{align*}
by Jensen's inequality. Using the Lipschitz assumptions on $\theta$ and $m$ combined with the definition of the spectral norm we have that the outer product satisfies
\begin{align*}
    ||\nabla_\alpha \theta_\alpha(x) [\nabla_\beta m(x, \beta)]^T||_2 \leq \kappa_\theta(x) \kappa_m(x)
\end{align*}
and hence
\begin{align*}
    ||\E[\nabla_\alpha \theta_\alpha(x) [\nabla_\beta m(x, \beta)]^T]||_2 
    \leq \E[\kappa_\theta(x) \kappa_m(x)].
\end{align*}
Since $H$ has bounded minimum eigenvalue, $H^{-1}$ has bounded spectral norm as well: $||H^{-1}||_2 \leq \frac{1}{\gamma}$. Finally, using the Lipschitz assumption on $h$ and again applying the sub-multiplicative property of the spectral norm, we have
\begin{align*}
    ||\nabla_\alpha f||_2 \leq \frac{1}{\gamma} \E[\kappa_\theta(x) \kappa_m(x)] \kappa_h.
\end{align*}
Stepping back to re-introduce the random function $\xi$, we have that the Lipschitz constant of $f$ with respect to $\alpha$ is $\frac{1}{\gamma_\xi} \E_\xi[\kappa_\theta(x) \kappa_m(x)] \kappa_h$. 

\textit{Step 2.}
Now we are going to proceed to prove that the restrictions are also Lipchitz. In order to do that, we are going to proceed using the same strategy as before and show that the restrictions also have bounded gradients. Let us denote the restrictions as $r_i(\alpha)$. Note that 
\begin{align*}
    \nabla_{\alpha}r_i = \frac{1}{n}g_i(X)\nabla_{\alpha} \theta(X) 
\end{align*}
Thus it must be true that:
\begin{align*}
    ||\nabla_\alpha r_i || = \frac{1}{n}|g_i(X)|||\nabla_{\alpha} \theta(X) ||
\end{align*}
As $g$ is continuous in a compact set it must be bounded by a finite number $M$. Then by hypothesis:
\begin{align*}
     ||\nabla_\alpha r_i|| \leq k_{\theta}(x)M
\end{align*}
And again by Jensen's inequality: 
\begin{align*}
     ||\nabla_\alpha r_i|| \leq \mathbb{E}[k_{\theta}(x)]M
\end{align*}

\textit{Step 3.} Since $f$ and all $r_i$ are Lipschitz on a finite-dimensional parameter, this implies that they belong to a Donsker class \cite{van2000asymptotic} and so we are guaranteed to a limiting stochastic process for each $r_i$ and for $f$ marginally. Furthermore, as $C(S)$ is separable, by theorem 3.2 of \cite{billingsley2013convergence}, the vector $(f,r_1,..,r_n)$ has a limiting process under the product measure.

\textit{Step 4.} Thus, as $\alpha \in S$ and because of the constraint qualification assumed for the problem, we may apply Theorem 3.6 of \cite{shapiro1991asymptotic} to guarantee Hadamard differentiability of the outer problem $\min_\alpha f(\alpha, \xi)$. Hence, the delta method in Theorem 2.1 of  \cite{shapiro1991asymptotic} can be applied to guarantee asymptotic normality and thus the result follows. 

\textit{Step 5} In order to explicitly compute the variance of $\sqrt{N}(\hat{\nu}_N - \nu)$, let
\begin{align*}
Y_n = \begin{bmatrix}
    h(\hat{\beta}_N(\alpha)) \\
    \frac{1}{n}\sum_i \theta_{\alpha}(X_i)g_1(X_i) \\
    \vdots \\
    \frac{1}{n}\sum_i \theta_{\alpha}(X_i)g_n(X_i)
\end{bmatrix}
\end{align*} and
\begin{align*}
\mu = \begin{bmatrix}
     h(\beta^*(\alpha)) \\
     \mathbb{E}[\theta_{\alpha}(X)g_1(X)] \\
    \vdots \\
    \mathbb{E}[\theta_{\alpha}(X)g_m(X)].
\end{bmatrix}
\end{align*}
Then, if $g(f) = \min_{\alpha \in \Theta}f(\alpha)$, again by theorem 2.1 \cite{shapiro1991asymptotic}
\begin{align*}
    \sqrt{n}(g(Y_n) - g(\mu)) \to_{D} g'_{\mu}(Z)
\end{align*}
Where $Z$ is the limiting object that we proved the empirical process has. As a consequence of theorem 3.6 of \cite{shapiro1991asymptotic}, if the set of minimizers is a singleton, $g'_{\mu}(\sqrt{n}(Y_n - \mu)) = \sqrt{n}[ h(\hat{\beta}_n(\alpha_0)) - h(\beta^*(\alpha_0))] + \sqrt{n}\sum_{j=1}^m\lambda_j[\frac{1}{n}\sum_i\theta_{\alpha_0}(X_i)g_j(X_i) - \mathbb{E}_{\QQ}[\theta_{\alpha_0}(X)g_j(x)]]$. The term $[h(\beta^*(\alpha_0)) -  h(\hat{\beta}_n(\alpha_0))]$ is normal as it is a M-estimator itself. Hence  $g'_{\mu}(\sqrt{n}(Y_n - \mu))$ is the sum of 
Normally distributed random variables and therefore,
\begin{align*}
    \sqrt{n}(\nu_n - \nu) \to N\left(0, Var\left(\sqrt{n}\left(h(\beta^*(\alpha_0)) -  h(\hat{\beta}_n(\alpha_0)) - \sum_{j=1}^m\lambda_j\frac{1}{n}\left[\sum_{i=1}^n\theta_{\alpha_0}(X_i)g_j(X_i)\right]\right)\right)\right).
\end{align*}
Using sample splitting and the delta method on $h(\beta^*)$, the variance can be approximated by $\nabla h(\beta^*(\alpha_0))^T \Sigma_{0} \nabla h(\beta^*(\alpha_0)) + Var(\sum_{i=j}^m\lambda^*_j \theta_{\alpha_0}(x)g_j(x))$, completing the proof.

\section{Experiments}
\subsection{Constraint derivation}
In some cases, we may be able to replace the constrain $\theta(X) \geq 0$ with a tighter constraint. For example, consider the case of \textit{selection bias}, where individual samples from $\PP$ select into $\QQ$ based on $X$. Formally, we model this via an indicator variable $R$, which is 1 if a unit is observed in the sample and 0 otherwise. Then, $q(x) = p(x|R = 1)$, and via Bayes theorem we have $\theta_0(X) = \frac{\text{Pr}(R = 1)}{\text{Pr}(R = 1|X)}$. Since $\text{Pr}(R = 1|X) \leq 1$, we are guaranteed that  $\theta_0(X) \geq \text{Pr}(R = 1)$. In many cases, the marginal $\text{Pr}(R = 1)$ is easily observable because we know the total size of the population that appears in our sample relative to the true population (e.g., a government may know the fraction of people in a city who are enrolled in a program). This allows us to replace the constraint $\theta(X) \geq 0$ with the tighter constraint $\theta(X) \geq \text{Pr}(R = 1)$. This was indeed the case in all the experiments presented in the article.

\subsection{Synthetic data experiments details}
 Inspired by data models used in the causal inference literature \cite{kennedy2019estimating}, the distribution  $X = (Y,Y_2, A, X_1, X_2) \sim \PP$ is given by the following model:
\begin{align*}
    X_1 &\sim Multinomial(3, 0.5, 0.3)\\
    X_2 &\sim Ber(0.4)\\
    A &\sim Ber(logit^{-1}(X_2 - X_1))\\
    Y &\sim Ber(logit^{-1}(2A - X_1 + X_2))\\
    Y_2 &\sim Ber(logit^{-1}((X_1 + X_2)/2 - A)
\end{align*}
 The observed distribution $\QQ$ is given by simulating selection bias via an indicator variable:
\begin{align*}
     R &\sim Ber(logit^{-1}(X_1 - X_2))
\end{align*}
Naturally, the sample from $\QQ$ are all those samples for which $R=1$. The set $\Theta$ used in the experiments was:
\begin{itemize}
    \item\textbf{Unrestricted}, \textbf{separable} and  \textbf{Targeted} : $ S:= \{\E_{X \sim \PP}  [Y_2X_2]= c_1, \E_{X \sim \PP}  [X_2(1 - Y_2)]= c_2, \E_{X \sim \PP}  [(1 - X_2)Y_2]= c_3, \E_{X \sim \PP}  [(1 - X_2)(1 - Y_2)]= c_4 \}$.
    \item \textbf{(partial) race + income :}, $\Theta$ was $S \setminus \{\E_{X \sim \PP}  [Y_2X_2]= c_1, \E_{X \sim \PP}  [(X_2 - 1)Y_2]= c_2\}$.
    \item \textbf{(full) race + income :} $\Theta$ was $S$.
    \item \textbf{race + income + Outcome :}  $\Theta$ was $S \cup \{\E_{X \sim \PP}  [YX_2]= c_5, \E_{X \sim \PP}  [Y(1 - X_2)]= c_6\}$.
\end{itemize}
 Each experiment was run 5 times over different bootstrap replicates of the sample of distribution $\PP$. The experimental results are summarized in table \ref{results-synthetic}.

\begin{table}[!ht]
  \caption{Synthetic data experimental results. True conditional mean: 0.733.}
  \label{results-synthetic}
  \centering
  \begin{tabular}{lccccc}
    \toprule
    \multicolumn{1}{c}{} & \multicolumn{1}{c}{} & \multicolumn{2}{c}{Lower bound} & \multicolumn{2}{c}{Upper bound} \\
    \cmidrule(r){3-4}
    \cmidrule(r){5-6}
    && mean & std & mean & std \\
    \midrule
    Separable & DRO & 0.475 & 0.004 & 0.814 & 0.003 \\
              & Ours & 0.671 & 0.016 & 0.786 & 0.033 \\
    Targeted & DRO & 0.501 & 0.004 & 0.800 & 0.004 \\
             & Ours & 0.697 & 0.003 & 0.816 & 0.002 \\
    Unrestricted & DRO & 0.403 & 0.004 & 0.844 & 0.002 \\
                 & Ours & 0.420 & 0.008 & 0.855 & 0.002 \\
    (Full) Race + Income & DRO & 0.406 & 0.002 & 0.842 & 0.002 \\
                            & Ours & 0.665 & 0.009 & 0.743 & 0.016 \\
                         
    (Partial) Race + Income & DRO & 0.405 & 0.002 & 0.843 & 0.002 \\
                            & Ours & 0.449 & 0.052 & 0.761 & 0.011 \\
    Race + Income + Outcome & DRO & 0.406 & 0.002 & 0.842 & 0.002 \\
                           & Ours & 0.665 & 0.008 & 0.744 & 0.014 \\
    DRO Omniscient &&	0.484	&0.004&	0.788&	0.002\\
    \bottomrule
  \end{tabular}
\end{table}

\subsection{Semi-synthetic data}

For the semi-synthetic experiments we used the Folkstables package \cite{ding2021retiring} which provides an interface for curated US Census data. We use the ACSEmployment task, where $Y$ is whether or not a person is employed, and the variable of interest $A$ is the sex of an individual. The rest of the covariates 
come from a one-hot encode of the features listed below. The last level of each variable is dropped to avoid an unidentifiable model, thus obtaining a 15-dimensional representation for every entry in the dataset.

\begin{enumerate}
    \item \textbf{Citizenship status}: 0: Born in the U.S, 1: Born in Puerto Rico, Guam, the U.S. Virgin Islands, or the Northern Marianas, 2: Born abroad of American parent(s), 3: U.S. citizen by naturalization, 4: Not a citizen of the U.S.
    \item \textbf{Military service}: 0: Is or was in active duty, 1: Never served in the military.
    \item \textbf{Nativity}:   0: Native, 1: Foreign-born.
    \item \textbf{disability}: 0: Not having any disability, 1: Having a Hearing, vision, or cognitive disability.
    \item \textbf{Income}: 0: Personal income over 50000 USD a year , 1: Personal income below 50000 USD a year.
    \item \textbf{Race}: 0:self-identifying as not white , 1:self-identifying as white
\end{enumerate}

 The observed distribution $\QQ$ is given by simulating selection bias via an indicator variable: 
\begin{align*}
     R &\sim Ber(logit^{-1}(X_1 - X_2))
\end{align*}
Again, the sample from $\QQ$ are all those samples for which $R=1$.

Let $Y$ be unemployment status, $A$ sex, $Y_2$ income and  $X_2$ be race. The set $\Theta$ used in the first setting of experiments (estimating $\E_{\PP}[Y|A = 1]$) is:
\begin{itemize}
    \item\textbf{Unrestricted}, \textbf{separable} and  \textbf{Targeted} : $ S:= \{\E_{X \sim \PP}  [Y_2X_2]= c_1, \E_{X \sim \PP}  [X_2(1 - Y_2)]= c_2, \E_{X \sim \PP}  [(1 - X_2)Y_2]= c_3, \E_{X \sim \PP}  [(1 - X_2)(1 - Y_2)]= c_4 \}$.
    \item \textbf{(partial) race + income :} $\Theta$ was $S \setminus \{\E_{X \sim \PP}  [Y_2X_2]= c_1, \E_{X \sim \PP}  [(X_2 - 1)Y_2]= c_2\}$.
    \item \textbf{(full) race + income :} $\Theta$ was $S$.
    \item \textbf{race + income + Outcome :}  $\Theta$ was $S \cup \{\E_{X \sim \PP}  [YX_2]= c_5, \E_{X \sim \PP}  [Y(1 - X_2)]= c_6\}$.
\end{itemize}

 Each experiment was run 5 times over different bootstrap replicates of the sample of distribution $\PP$. The experimental results are summarized in table \ref{results-semi-synthetic}.

\begin{table}[!ht]

  \caption{Semi-synthetic data experimental results. True conditional mean: 0.376.}
  \label{results-semi-synthetic}
  \centering
  \begin{tabular}{lccccc}
    \toprule
     \multicolumn{1}{c}{} & \multicolumn{1}{c}{} & \multicolumn{2}{c}{Lower bound} &  \multicolumn{2}{c}{Upper bound} \\
    \cmidrule(r){3-4}
    \cmidrule(r){5-6}
    && mean     & std & mean & std\\
    \midrule
    Unrestricted & DRO & 0.132 & 0.055 & 0.602 & 0.107 \\
                 & Ours & 0.161 & 0.003 & 0.509 & 0.004 \\
    Separable & DRO & 0.129 & 0.004 & 0.433 & 0.008 \\
              & Ours & 0.371 & 0.007 & 0.383 & 0.009 \\
    Targeted & DRO & 0.129 & 0.004 & 0.433 & 0.008 \\
             & Ours & 0.366 & 0.015 & 0.385 & 0.010 \\
    (Partial) Race + Income & ours & 0.067	& 0.004	& 0.666	& 0.012\\
    	                    & DRO  & 0.064 &	0.004   &	0.753	& 0.015 \\
    (Full) Race + Income    & ours & 0.160 &	0.004	&   0.513	& 0.006 \\
                            &DRO   & 0.093 &	0.003	&   0.678   & 0.014\\
     Race + Income + Outcome&ours  & 0.184 &    0.008   &   0.389   & 0.007\\
                            &DRO	&0.093&	0.003&	0.677&	0.013\\
    DRO Omniscient	&& 0.208 &	0.012	&0.451	&0.01 \\
    \bottomrule
  \end{tabular}
\end{table}

The set $\Theta$ used in the second setting of experiments (estimating the coefficient $\beta$ of the indicator variable $A$ in a linear regression model) is:
\begin{itemize}
    \item\textbf{Unrestricted} and \textbf{targeted}: $ S:= \{\E_{X \sim \PP}  [Y_2X_2]= c_1, \E_{X \sim \PP}  [X_2(1 - Y_2)]= c_2, \E_{X \sim \PP}  [(1 - X_2)Y_2]= c_3, \E_{X \sim \PP}  [(1 - X_2)(1 - Y_2)]= c_4 \}$.
    \item \textbf{(Full) Race + income :}, $\Theta$ was $S \setminus \{\E_{X \sim \PP}  [Y_2X_2]= c_1, \E_{X \sim \PP}  [Y_2(X_2-1)]= c_2\}$.
    \item \textbf{Race + income + Outcome : } $\Theta$ was $S$.
\end{itemize}
Each experiment was run 5 times over different bootstrap replicates of the sample of distribution $\PP$. The experimental results are summarized in table \ref{results-regression}.

\begin{table}[!ht]
  \caption{Semi-synthetic data experimental results. True conditional mean: -0.305.}
  \label{results-regression}
  \centering
  \begin{tabular}{lccccc}
    \toprule
    \multicolumn{1}{c}{} & \multicolumn{1}{c}{} & \multicolumn{2}{c}{Lower bound} & \multicolumn{2}{c}{Upper bound} \\
    \cmidrule(r){3-4}
    \cmidrule(r){5-6}
    && mean & std & mean & std \\
    \midrule
    Unrestricted & DRO & -0.513 & 0.026 & 0.112 & 0.011 \\
                 & Ours & -0.492 & 0.022 & 0.062 & 0.036 \\
    Separable & DRO & -0.345 & 0.013 & -0.055 & 0.033 \\
              & Ours & -0.305 & 0.010 & -0.303 & 0.009 \\
    
    (Full) Race + Income & DRO & -0.492 & 0.022 & 0.062 & 0.036 \\
                         & Ours & -0.515 & 0.028 & 0.123 & 0.025 \\
    Race + Income + Outcome & DRO & -0.509 & 0.029 & 0.110 & 0.017 \\
                           & Ours & -0.304 & 0.057 & -0.079 & 0.011 \\
    \bottomrule
  \end{tabular}
\end{table}

\subsection{Logistic regression model}

\begin{figure}[!ht]
  \centering
  
  \includegraphics[scale=0.4]{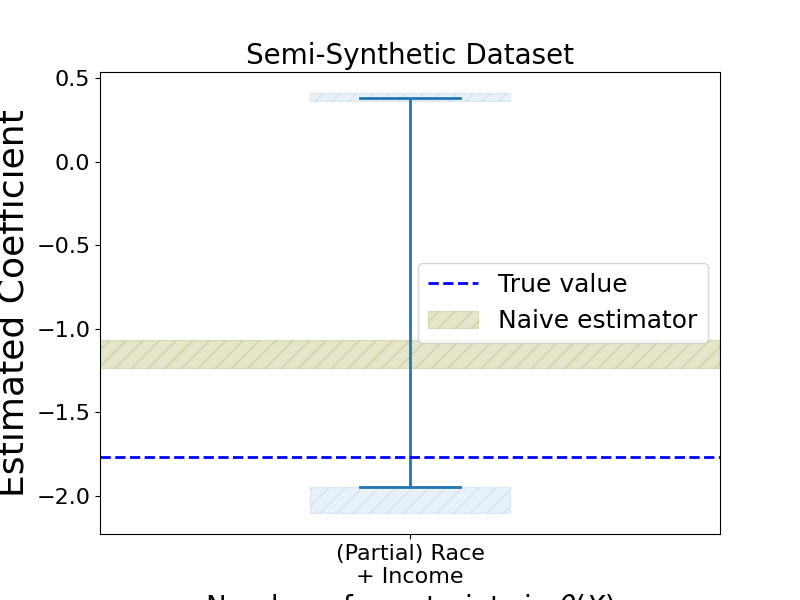}
  \caption{Bounds outputted by our method for the coefficient $\beta$ of the indicator variable $A$ .}

  \label{bounds_beta_logisitic}
\end{figure}

We run one additional experiment to estimate the coefficient $\beta$ of the indicator variable $A$. However, instead of being the coefficient of a linear model, it is now the coefficient of a logistic regression. Only the \textbf{unrestricted} experiment was run. The experiment was run 5 times. The data reported is the average value and standard deviation for the 5 outputs obtained from the experiment. The results are summarized in figure \ref{bounds_beta_logisitic} and table \ref{results-semi-synthetic-beta_logistic}.

\begin{table}[!ht]
  \caption{Semi-Synthetic data experimental results. True conditional mean: -1.769.}
  \label{results-semi-synthetic-beta_logistic}
  \centering
  \begin{tabular}{lcccc}
    \toprule
    \multicolumn{1}{c}{} & \multicolumn{2}{c}{Lower bound} & \multicolumn{2}{c}{Upper bound} \\
    \cmidrule(r){2-3}
    \cmidrule(r){4-5}
    & mean & std & mean & std \\
    \midrule
    Unrestricted & -1.998 & 0.069 & 0.382 & 0.019 \\
    \bottomrule
  \end{tabular}
\end{table}

\subsection{Covariance-like restrictions}

We run one additional experiment to estimate the coefficient $\beta$ of the indicator variable $A$. However, instead of being the coefficient of a linear model, it is now the coefficient of a logistic regression. The experiment was run 5 times. The data reported is the average value and standard deviation for the 5 outputs obtained from the experiment. The results are in figure \ref{bounds_beta_logisitic} and table \ref{results-semi-synthetic-beta_logistic}. For the results in figure \ref{bounds_beta_logisitic} and table \ref{results-semi-synthetic-beta_logistic}, we constrained the covariance for a set of variables to be positive. Specifically, we took the pair of variables with the largest positive covariance\textemdash being a US citizen and being a native\textemdash and restricted that covariance to positive during the optimization.

Overall, we experimented with two types of restrictions on the covariance: restricting the sign of the covariance or the exact covariance value. The two types of constraints had similar results. We also tested restricting on multiple feature pairs but didn’t see any significant differences in the bounds.

\begin{figure}[!ht]
  \centering
  
  \includegraphics[scale=0.4]{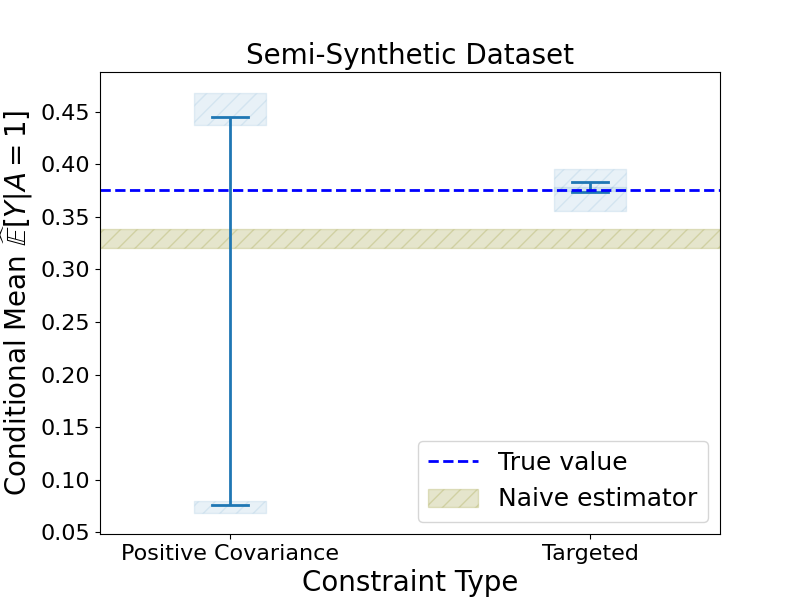}
  \caption{Bounds outputted by our method for the coefficient $\beta$ of the indicator variable $A$ .}

  \label{covariance_bounds}
\end{figure}

\subsection{Guide to run the experiments}

In the \textbf{experiments\_metadata} folder are JSON files with the exact hyperparameters used in all the experiments presented in the paper. To execute a particular set of experiments related to the conditional mean, use the \texttt{inference.py} script. For example, the following command generates the bounds with different parametric forms for the simulated dataset:
\begin{lstlisting}[language=sh, caption={Run Python Script}, label=python-script]
python inference.py experiments_metadata/sim_theta_ours.json
\end{lstlisting}

To recreate the results of the experiments that involve regression models, use the \texttt{inference\_non\_closed\_form.py} script: 
\begin{lstlisting}[language=sh, caption={Run Python Script}, label=python-script]
python inference_non_closed_form.py experiments_metadata/non_closed_logistic.json
\end{lstlisting}

We will provide a detailed table with the JSON files that generate the data of the plots we included.

Each successful run of the two commands will produce a folder containing a data frame with the results and several plots. The folder's name is a timestamp denoting the experiment's end time. To reproduce the plots from the paper, use the notebook in the folder named \textbf{plotting}.

The directory hierarchy is the following
\dirtree{%
.1 / (root).
.2 README.md.
.2 experiment\_artifacts.
.2 optimization.
.3 experiments\_metadata.
.3 inference.py.
.3 inference\_non\_closed\_form.py.
.2 plotting.
.3 paper\_plots.ipynb.
.3 utils.py.
}

\end{document}